\documentclass{article}

\usepackage[dblblindworkshop, final]{neurips_2025}

\usepackage[utf8]{inputenc} % allow utf-8 input
\usepackage[T1]{fontenc}    % use 8-bit T1 fonts
\usepackage{hyperref}       % hyperlinks
\usepackage{url}            % simple URL typesetting
\usepackage{booktabs}       % professional-quality tables
\usepackage{amsfonts}       % blackboard math symbols
\usepackage{nicefrac}       % compact symbols for 1/2, etc.
\usepackage{microtype}      % microtypography
\usepackage{xcolor}         % colors
\usepackage{colortbl}
\usepackage{amsmath}       % colors
\usepackage{comment}
\usepackage{graphicx}
\usepackage[table]{xcolor}
\usepackage{multirow}
\definecolor{cLow}{HTML}{67001F}
\definecolor{cMid}{HTML}{F4A582}
\definecolor{cHigh}{HTML}{FFFFBF}
\definecolor{orangeLow}{RGB}{255, 224, 178}   % light orange
\definecolor{orangeMid}{RGB}{255, 152, 0}     % medium orange
\definecolor{orangeHigh}{RGB}{230, 81, 0}     % dark orange
\title{Robust Neural Audio Fingerprinting \\using Music Foundation Models}

\author{
\begin{minipage}{\textwidth}
\centering
\textbf{Shubhr Singh}$^{1}$\thanks{Equal contribution.}\quad
\textbf{Kiran Bhat}$^{1}$\footnotemark[1]\quad
\textbf{Xavier Riley}$^{1}$\quad
\textbf{Benjamin Resnick}$^{1}$\\[0.4em]
\textbf{John Thickstun}$^{1,2}$\quad
\textbf{Walter De Brouwer}$^{1}$\\[0.75em]
$^{1}$SoundPatrol \quad $^{2}$Cornell University
\end{minipage}
}
\begin{document}

\maketitle

\begin{abstract}
The proliferation of distorted, compressed, and manipulated music on modern media platforms like TikTok motivates the development of more robust audio fingerprinting techniques to identify the sources of musical recordings.
In this paper, we develop and evaluate new neural audio fingerprinting techniques with the aim of improving their robustness. We make two contributions to neural fingerprinting methodology: (1) we use a pretrained music foundation model as the backbone of the neural architecture and (2) we expand the use of data augmentation to train fingerprinting models under a wide variety of audio manipulations, including time streching, pitch modulation, compression, and filtering.
We systematically evaluate our methods in comparison to two state-of-the-art neural fingerprinting models: NAFP and GraFPrint. Results show that fingerprints extracted with music foundation models (e.g., MuQ, MERT) consistently outperform models trained from scratch or pretrained on non-musical audio. Segment-level evaluation further reveals their capability to accurately localize fingerprint matches, an important practical feature for catalog management.
\end{abstract}

\section{Introduction}
Audio fingerprinting  identifies unknown audio by extracting compact feature representations, or \emph{fingerprints}, from a query and matching them against a reference database~\cite{oguz2025enhancing}. Fingerprinting has a wide range of applications such as music identification \cite{shazam}, integrity verification~\cite{gomez2002mixed}, and broadcast monitoring~\cite{cortes2022baf}. Queries often differ from reference tracks due to environmental degradations (e.g., noise, reverberation, microphone coloration) or deliberate modifications (e.g., pitch shifts, time stretches, lossy compression). Effective fingerprints should be both robust to such variations and discriminative enough to distinguish tracks.

Recent progress in neural audio fingerprinting~\cite{nafp} has shifted the field beyond classical methods like Shazam~\cite{shazam}, towards contrastive learning~\cite{simclr} approaches that align representations of original and modified audio~\cite{nafp,grafprint}. Prior approaches to neural audio  fingerprinting either learn these representations from scratch~\cite{nafp, grafprint} or adapt them from general-purpose audio models~\cite{magcil}. In this work we extend this line of research by exploring  music foundation models as pretrained backbones for fingerprinting. Furthermore, we systematically evaluate each model's robustness against a broader set of manipulations that characterize modern media ecosystems, extending beyond noise and reverberation to include time steching~\cite{umg2024believe}, pitch shifting, compression, and filtering. 

We focus on fingerprints derived from two music foundation models,  MuQ~\cite{zhu2025muq} and MERT~\cite{li2024mert}, as well as the general-purpose audio foundation model BEATs~\cite{chen2022beats}, previously considered for fingerprinting by~\cite{magcil}. To contextualize the performance of these fingerprints, we benchmark against two state of the art neural fingerprinting models, NAFP~\cite{nafp} and GraFPrint~\cite{grafprint}, as well as a Shazam-like baseline implemented with open source library Dejavu~\cite{dejavu}. All models are assessed under a broad collection of audio distortions and manipulations, for both track-level and segment-level identification tasks.

% We systematically evaluate two state-of-the-art neural fingerprinting models, NAFP~\cite{nafp} and GraFPrint~\cite{grafprint}, together with a Shazam-like audio fingerprinting approach implemented through the open-source library Dejavu~\cite{dejavu}. We compare them against fingerprints based on the self-supervised music foundation models MuQ~\cite{zhu2025muq} and MERT~\cite{li2024mert}, as well as the general-purpose audio foundation model BEATS~\cite{chen2022beats}, across a variety of audio modification techniques on both track-level and segment-level tasks.%  Our results show that the pretrained backbones consistently demonstrate greater robustness than their counterparts trained from scratch. All models are trained on four NVIDIA H100

Previous work on neural audio fingerprinting focuses on track-level evaluation with the Free Music Archive (FMA) dataset~\cite{fma_dataset}, using random splits of the same dataset for both training and testing. This setup offers only a limited view of real-world deployment, where query and reference databases can come from different distributions. To address this, we train and evaluate fingerprints on separate datasets to assess a fingerprinting model's robustness to subtle shifts in the data distributions. Fingerprinting experiments are conducted for two retrieval tasks: traditional \textit{track-based} identification~\citep{nafp} and additionally \textit{segment-based} identification, following the Pexeso Benchmark~\cite{pexeso_audio},a standardized open-source framework for evaluating fingerprint retrieval and temporal alignment under controlled degradations.

% We train and evaluate the models on separate datasets, in contrast to prior studies that primarily relied on training and testing with different subsets of the Free Music Archive (FMA) dataset~\cite{fma_dataset}. By doing so, we consider deployment scenarios where the reference database and the query audio often originate from different distributions. In line with this goal of generalizability, we also reconsider existing evaluation methodologies, which have primarly emphasized track level matching to compute Top-1 hit rats~\cite{nafp}. While useful, such metrics overlook temporal correspondance between query and reference set. We therefore extend the evaluation to segment level boundary detection in line with the Pexeso Benchmark~\cite{pexeso_audio}, reporting: \textbf{track-level F1} (correct file-pair retrieval), \textbf{length-level F1} (fraction of query duration correctly matched), and \textbf{bounding-box F1} (segment boundary accuracy using the copy-overlap protocol~\cite{he2022large}).

% Evaluation protocols have also focused primarily on track-level matching to compute Top-1 accuracy~\cite{nafp,grafprint}, providing no insight into the precise temporal alignment between query and reference. Incorporating segment-level boundary detection could offer valuable information about the exact position within a track, which is particularly important when operating over large-scale catalogs. 

\begin{figure}[t]
    \hspace*{-1cm} % shift left
    \includegraphics[width=1.1\linewidth]{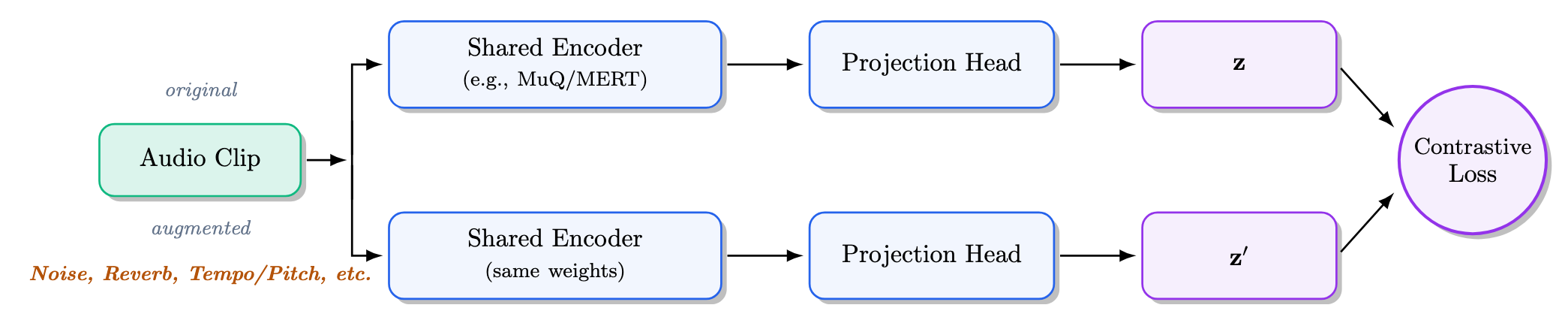}
    \caption{The contrastive learning framework for neural audio fingerprinting. Original and augmented audio (e.g., audio with noise, reverb, time/pitch changes) are passed through a shared encoder, followed by a projection head. The resulting embeddings ($z$ and $z'$) are optimized using a contrastive loss to encourage invariance to audio degradations.}
    \label{fig:methodology}
\end{figure}
\section{Methodology}
This section describes our methodology for fingerprinting: datasets (Section~\ref{sec:datasets}), fingerprint representation learning (Section~\ref{sec:models}), data augmentations (Section~\ref{subsec:augmentations}), and inference algorithms (Section~\ref{sec:inference}).

% experimental setup used to evaluate the proposed approach. 
% The experiments span two distinct retrieval settings: \textit{track-based} and \textit{segment-based}. 
% In the track-based setup, queries correspond to modified excerpts of complete reference tracks, while in the segment-based setup, queries contain multiple short segments sourced from different references, potentially under severe degradations. 

\subsection{Datasets}\label{sec:datasets}
For training, we use 300{,}000 samples from the Disco-10M dataset~\cite{lanzendorfer2023disco}, ensuring no overlap with the FMA dataset~\cite{fma_dataset}. For the \textit{track-based} evaluation, 5{,}000 reference tracks are drawn from FMA, with 1{,}000 of these used to generate query files by extracting random 10-second segments and applying a range of audio distortions (see Section~\ref{subsec:augmentations}). For the \textit{segment-based} evaluation, we follow the Pexeso Audio Fingerprinting Benchmark Toolkit~\cite{pexeso_audio} using the \texttt{pexafb\_hard\_small} and \texttt{pexafb\_hard\_medium} difficulty levels, which provide reference sets of 99 and 953 files and corresponding query sets of 100 and 1{,}000 files. 
In this setting, queries are constructed by concatenating one or more 10-second segments, each modified with distortions such as time streching, pitch shifts, echo, reverb, filtering, or noise, and concatenated using techniques such as fades or overlaps. 

\subsection{Models}\label{sec:models}
We evaluate a unified fingerprinting architecture consisting of a pretrained encoder followed by a non-linear projection head. Given an embedding \(\mathbf{x}\in\mathbb{R}^{d_{\text{in}}}\) from the backbone (\(d_{\text{in}}=1024\)), we  project \(\mathbf{x}\) to \(\mathbf{z}\in\mathbb{R}^{d_{\text{out}}}\) (\(d_{\text{out}}=256\)) using a two-layer MLP (Projection Head in Fig ~\ref{fig:methodology}).

\[
\mathbf{z} \;=\; \mathbf{W}_2\,\phi(\mathbf{W}_1 \mathbf{x} + \mathbf{b}_1) + \mathbf{b}_2,
\]
where \(\mathbf{W}_1 \in \mathbb{R}^{d_h \times d_{\text{in}}}\), \(\mathbf{b}_1 \in \mathbb{R}^{d_h}\), \(\mathbf{W}_2 \in \mathbb{R}^{d_{\text{out}} \times d_h}\), \(\mathbf{b}_2 \in \mathbb{R}^{d_{\text{out}}}\), \(d_h=4096\), and \(\phi\) is ELU~\cite{clevert2015fast}  nonlinearity. In early ablations, we found that this two-layer projection head with non-linearity outperforms the linear projection head used in prior works such as NAFP and GraFPrint; use of a large hidden width $d_h$ further improves performance. 

For the backbone, we consider three foundation models. MuQ~\cite{zhu2025muq} is a self-supervised music representation model based on masked token prediction with Conformer blocks~\cite{gulati2020conformer}, designed to capture both local and long-range musical structure. MERT~\cite{li2024mert} adapts masked language modeling to musical audio using a convolutional front-end and Transformer encoder to jointly predict masked acoustic tokens. BEATs~\cite{chen2022beats} is a general-purpose audio foundation model pretrained on AudioSet-2M~\cite{gemmeke2017audio}. For MuQ and MERT, we set the learning rate to $3\times 10^{-5}$, while BEATs is trained with $5 \times 10^{-5}$. We compare these methods against NAFP~\cite{nafp}, which employs a convolutional backbone trained from scratch with contrastive learning, and GraFPrint~\cite{grafprint}, which extends NAFP with a graph neural network (also trained from scratch). Dejavu~\cite{dejavu} serves as a Shazam-like baseline that extracts constellation maps of spectral peaks, encodes them as landmark-based hashes, and retrieves matches using hash-table lookups.

\subsection{Data Augmentations}
\label{subsec:augmentations}
We employ a range of augmentations during both training and evaluation to improve robustness against common acoustic and signal-domain variations. Temporal modifications include \emph{time streching} (uniform in $[0.7, 1.5]$) and \emph{pitch-shifting} (uniform in $[-5, 5]$ semitones), applied individually or sequentially. Additive noise augmentation uses $\sim$6 hours of MUSAN~\cite{snyder2015musan} recordings (restaurant, home, street) at varying SNRs, optionally combined with reverberation simulated using RIRs from the Aachen database~\cite{jeub2009binaural} (RIR SNR in $[0.1, 1.5]$). Spectral filtering is applied using band-pass ($300$–$1800$ Hz), high-pass ($[1800, 3400]$ Hz), or low-pass ($[300, 1500]$ Hz) filters to emulate telephony, bandwidth-limited playback, or lo-fi effects. Echo is introduced with delays in $[100, 200]$ ms, while low-bitrate artifacts are simulated using Encodec~\cite{defossezhigh} at 6-bit and 12-bit quantization (24 kHz model). 

\subsection{Inference}\label{sec:inference}
For track-level retrieval, we follow a conventional inference procedure. Query and reference tracks are passed through the trained model to generate embeddings, which serve as audio fingerprints. Reference fingerprints form a database and queries are matched using FAISS~\cite{johnson2019billion},an efficient library for approximate nearest-neighbour search over large embedding databases.  A query is counted as correct if this retrieved reference matches the ground-truth reference for that query. 

% In order to compute the top-1 hit rate, the top-5 reference matches are retrieved for every query segment and then aggregated by majority vote over reference IDs to yield a single file-level reference prediction.
The segment-level retrieval task defined by the Pexeso benchmark is a novel setting for fingerprinting, where queries are constructed from multiple snippets originating from different reference tracks. This requires a new approach to inference that both identifies matches and localizes them with temporal alignment. To address this, we obtain the top-5 FAISS neighbors per query segment, filter out candidates below a fixed similarity threshold (0.7), and group the remaining matches by (query file, reference file). Within each group, every retained match gives two numbers: the segment’s start time in the query and the corresponding start time in the reference. We then fit a linear model \( t_{\mathrm{ref}} \approx a\,t_{\mathrm{qry}} + b \) using Huber regression~\cite{huber1992robust}, which down-weights outliers. The parameter \(a\) is a \emph{time-scaling factor} that captures a uniform speed discrepancy between the query and the reference. When \(a=1\), the two run at the same speed, while \(a>1\) indicates the query runs more slowly.

The inlier matches within each \((\text{query},\text{reference})\) group trace one or more candidate timing trajectories. For each group, we evaluate the trajectories generated from different seeds and keep the strongest one, giving priority to trajectories with more inliers and higher goodness of fit (larger \(R^{2}\)). 
The selected trajectory is then converted to segment boundaries by taking the earliest inlier as the start time  and the latest inlier plus segment length as the end time on both query and reference segment boundaries. The resulting interval is taken as the aligned match and is scored against the ground-truth annotations.

\section{Results \& Discussion}
% queries correspond to modified excerpts of reference tracks; for the segment-based task, queries consist of multiple shorter segments sourced from different reference tracks, potentially under severe degradations. For track-based retrieval,

% For the track-based retrieval,  fingerprint performance is measured using the top-1 hit rate, i.e., the percentage of queries whose correct reference track is ranked first.
% For segment-based retrieval, we follow the Pexeso Benchmark~\cite{pexeso_audio}, reporting: \textbf{track-level F1} (correct file-pair retrieval), \textbf{length-level F1} (fraction of query duration correctly matched), and \textbf{bounding-box F1} (segment boundary accuracy using the copy-overlap protocol~\cite{he2022large}). 

\renewcommand{\arraystretch}{1.2} % adjust row height
\setlength{\tabcolsep}{4pt} % reduce column spacing

\begin{table}[t!]
\caption{Top-1 hit rate (\%) on track-level evaluation. T+P denotes both time stretch and pitch shift applied, R+N denotes reverb and noise combinations, B.P. denotes band-pass filtering, H.P. denotes high-pass filtering, L.P. denotes low-pass filtering, and Enc. denotes Encodec compression.}
\label{tab:tab1}
\centering
\small
\begin{tabular}{l l l l l l l l l l l l l}
\hline
\textbf{Model} & \textbf{Time} & \textbf{Pitch} & \textbf{T+P} & \textbf{Noise} & \textbf{Reverb} & \textbf{R+N} & \textbf{B.P.} & \textbf{H.P.} & \textbf{L.P.} & \textbf{Echo} & \textbf{Enc.} & \textbf{Overall} \\
\hline
\textbf{MuQ-Unfrozen} & 96 & 94 & \textbf{87} & \textbf{97} & \textbf{100} & \textbf{90} & \textbf{63} & \textbf{73} & 74 & \textbf{100} & \textbf{96} & \textbf{88.18} \\
\textbf{MuQ-Frozen} & 90 & 91 & 86 & 90 & 98 & 84 & 60 & 72 & 69 & 93 & 90 & 83.91 \\
\textbf{MERT-Unfrozen} & \textbf{100} & 92 & 81 & 87 & 98 & 78 & 32 & 35 & 70 & \textbf{100} & 44 & 74.27 \\
\textbf{MERT-Frozen} & 97 & 89 & 81 & 86 & 95 & 71 & 30 & 29 & 68 & 96 & 38 & 70.91 \\
\textbf{BEATs-Unfrozen} & 84 & 89 & 73 & 84 & 91 & 77 & 27 & 39 & 76 & \textbf{100} & 33 & 70.27 \\

\textbf{GraFPrint} & 58 & \textbf{97} & 67 & 90 & 84 & 80 & 15 & 47 & 95 & 96 & 17 & 67.82 \\
\textbf{NAFP} & 39 & 91 & 55 & 84 & 86 & 78 & 18 & 42 & \textbf{96} & 99 & 10 & 63.45 \\
\textbf{Dejavu} & 25 & 91 & 12 & 71 & 80 & 52 & 9 & 12 & 87 & 99 & 3 & 49.18 \\
\hline
\end{tabular}
\end{table}

\renewcommand{\arraystretch}{1.1} % increase row height
\begin{table}[h]
\caption{Segment-level evaluation (F1 scores, \%) on Pexeso benchmark (hard settings). BBox refers to bounding-box.}
\label{tab:tab2}
\centering
\resizebox{0.9\textwidth}{!}{
\small
\begin{tabular}{l|ccc|ccc}
\hline
\multirow{2}{*}{\textbf{Model}} & \multicolumn{3}{c|}{\textbf{Pex-Hard-Small}} & \multicolumn{3}{c}{\textbf{Pex-Hard-Medium}} \\
 & \textbf{Track F1} & \textbf{BBox F1} & \textbf{Length F1} & \textbf{Track F1} & \textbf{BBox F1} & \textbf{Length F1} \\
\hline
\textbf{MuQ-Unfrozen}   & \textbf{95.5} & \textbf{86.4} & \textbf{90.8} & \textbf{87.3} & \textbf{74.7} & \textbf{81.4} \\
\textbf{MuQ-Frozen}     & 91.1 & 83.0 & 88.1 & 84.8 & 73.8 & 80.1 \\
\textbf{MERT-Unfrozen}  & 92.47 & 80.20 & 86.3 & 85.55 & 71.00 & 78.85 \\
\textbf{MERT-Frozen}    & 84.10 & 71.75 & 70.88 & 80.75 & 59.10 & 69.10 \\
\textbf{BEATs-Unfrozen}    & 80.8 & 73.10 & 76.4 & 85.2 & 70.2 & 76.70 \\
\textbf{GraFPrint}      & 78.00 & 49.40 & 67.60 & 81.30 & 61.90 & 70.40 \\
\textbf{NAFP}           & 77.47 & 40.50 & 66.41 & 80.6 & 57.4 & 66.8 \\

\textbf{Dejavu}           & 67.8 & 40.2 & 63.11 & 73.2 & 58.4 & 68.8 \\
\hline
\end{tabular}
}
\end{table}

% --------------------------

Track-level retrieval results on the curated FMA dataset highlight clear differences between pretrained backbones, neural baselines, and the Shazam-like system. As shown in Table~\ref{tab:tab1}, the pretrained backbones (MuQ, MERT, and BEATs) consistently surpass state-of-the-art models trained from scratch and classical methods, highlighting the value of pretraining. For the music foundation models (MuQ and MERT), we show two settings: frozen (encoder weights fixed) and unfrozen (encoder fine-tuned). In the case of BEATs, we evaluate with the unfrozen version. The unfrozen MuQ model achieves the highest overall accuracy, clearly outperforming all other models, and is particularly robust to Encodec compression, a setting where other models struggle. However, relative to their own performance across augmentations, both MuQ and MERT show reduced accuracy for filtering-based augmentations. Interestingly, the NAFP model outperforms all the other models in case of low pass filtering augmentation by a significant margin. The causes of this sensitivity will be investigated in future work. Across all conditions, neural approaches consistently surpass the Shazam-like baseline implemented with Dejavu.

 % While NAFP and GraFPrint perform reasonably well under noise and reverberation, they show weaker performance for tempo and combined tempo–pitch shifts, where the pretrained backbones excel.
Segment-level evaluation results on the Pexeso benchmark further confirm the contrast between the pretrained backbones and neural baselines. Table~\ref{tab:tab2} shows that the unfrozen MuQ model achieves the highest scores across all metrics, namely \textbf{track-level F1} (reference retrieval), \textbf{length-level F1} (alignment accuracy), and \textbf{bounding-box F1} (segment boundary precision~\cite{he2022large}).Consistent with the track-level results, MuQ, MERT, and BEATs outperform NAFP and GraFPrint by a clear margin. %Interestingly however, NAFP and GraFPrint improve when moving from the smaller to the medium-sized dataset, while the pretrained backbones exhibit a slight decline. This divergent trend raises questions about scalability, specifically whether the performance margins compared to the pretrained backbones would narrow in large-scale fingerprinting scenarios; a full exploration of this effect is left for future work. 
Interestingly, NAFP and GraFPrint improve from the smaller to the medium-sized dataset, while pretrained backbones slightly decline, highlighting a scalability contrast that warrants further investigation in large-scale fingerprinting scenarios
\section{Conclusion}
This work presents a systematic evaluation of self-supervised music foundation models (MuQ, MERT) and a general-purpose audio foundation model (BEATs) against state-of-the-art neural audio fingerprinting approaches (NAFP, GraFPrint), under a broad set of audio modifications at both track and segment levels. Models with pretrained backbones consistently outperform those trained from scratch, showing superior robustness and generalization, especially under challenging conditions. Segment-level evaluation further highlights their ability to accuratsly localize matched regions, an important capability for large-scale catalog management.
Our findings suggest that pretrained music foundation models can serve as powerful backbones for audio fingerprinting, but also reveal weaknesses in handling certain transformations, such as spectral filtering. Future work will explore targeted augmentation strategies to address these weaknesses and extend the evaluation to include new types of adversarial audio changes that are intentionally used to avoid detection on modern content-sharing platforms.

\bibliographystyle{unsrt}
\bibliography{refs}
\end{document}